\documentclass[a4paper,11pt]{article}
\usepackage{pos,placeins}

\title{The Roberge-Weiss endpoint in $(2+1)$-flavor QCD with background magnetic fields}
\author*[a]{K. Zambello}
\author[a]{M. D'Elia}
\author[a,b]{L. Maio}
\author[a]{G. Zanichelli}

\affiliation[a]{Università di Pisa and INFN, Sezione di Pisa, Pisa, Largo B. Pontecorvo 3, I-56127 Pisa, Italy}

\affiliation[b]{Aix Marseille Univ, Universit\'{e} de Toulon, CNRS, CPT, Marseille, France}

\emailAdd{kevin.zambello@pi.infn.it}
\emailAdd{massimo.delia@unipi.it}
\emailAdd{lorenzo.maio@roma2.infn.it}
\emailAdd{giuseppe.zanichelli@studenti.unipi.it}

\abstract{In this work we discuss our preliminary results regarding the so-called Roberge-Weiss (RW) transition, which is
  found for imaginary values of the baryon chemical potential, in the presence of a background magnetic field.
  We perform lattice QCD simulations on $N_t=6,8$ lattices with $2+1$ flavors of stout-staggered fermions at physical quark masses and the tree-level Symanzik improved gauge action. We determine the location the RW endpoint at finite magnetic fields and we study the order of the transition.}

\FullConference{The 41st International Symposium on Lattice Field Theory (LATTICE2024)\\
 28 July - 3 August 2024\\
Liverpool, UK\\}

\begin{document}
\maketitle

\section{Introduction}

The QCD phase diagram at finite temperature and density is extensively studied due to its importance for understanding various physical phenomena,
such as heavy-ion collision experiments and compact stars. At zero chemical potential the confined hadronic phase is known to undergo a transition
to the deconfined quark-gluon plasma phase. It is a true phase transition only in the vanishing / infinite quark masses limits, where it is
associated with the spontaneous symmetry breaking of the chiral / center symmetry. For physical quark masses the transition is a crossover
taking place at $T_c \simeq 155$~MeV~\cite{Aoki:2006we,Aoki:2006br,Borsanyi:2010bp,Bazavov:2011nk,Bhattacharya:2014ara},
but it is expected to become a true first order phase transition at high chemical potentials, starting from a critical point where the transition
is second order. Unfortunately, first-principles calculations in this regime are hindered by the sign problem. Despite this challenge, considerable
theoretical and experimental efforts are being made to locate the critical endpoint.

From a theoretical perspective, QCD at finite density can be studied, for instance, through simulations at purely imaginary chemical potentials. Since the theory at imaginary $\mu_q$ is free of the sign problem, lattice simulations can be conducted and numerical results can be extrapolated to real chemical potentials by analytic continuation~\cite{Alford:1998sd,Lombardo:1999cz,deForcrand:2002hgr,DElia:2002tig}. At imaginary $\mu_q$ the theory also has a phase structure interesting on its own. It exhibits a new symmetry, the Roberge-Weiss symmetry~\cite{Roberge:1986mm}. This is a remnant of the center symmetry and is related to the presence of first order phase transition lines between phases characterized by a different orientation of the Polyakov loop, at temperatures greater than the Roberge-Weiss temperature $T_{RW}$, above which the symmetry is spontaneously broken. The Roberge-Weiss lines lie on the chemical potentials $\mu_q / T = i ~ (2k + 1) ~ \pi / 3$ and they end up at critical endpoints (the RW endpoints) which {have been studied extensively in the
  literature~\cite{DElia:2009bzj,deForcrand:2010he,Bonati:2010gi,Cea:2012ev,Wu:2013bfa,Philipsen:2014rpa,Wu:2014lsa,Nagata:2014fra,Kashiwa:2016vrl,Makiyama:2015uwa,Czaban:2015sas,Philipsen:2019ouy}}
 and are thought to be second order for physical quark masses~\cite{Bonati:2016pwz,Dimopoulos:2021vrk}.

 The phase diagram can also be extended by introducing other phenomenologically relevant parameters. In particular since strong magnetic fields arise
 in non-central heavy-ion collisions and could have played a role in the evolution of the early Universe, one can introduce an external magnetic
 field and investigate the effects of this parameter on theoretical predictions (for a recent review, see Ref.~\cite{Endrodi:2024cqn}).
 Previous research showed that in the presence
 of a magnetic field the chiral restoration temperature decreases~\cite{Bali:2011qj}. Additionally, chiral symmetry breaking is enhanced at zero temperature,
 even though the opposite happens at higher temperatures, an effect known as inverse catalysis~\cite{Bali:2012zg} which however depends on the pion
 mass~\cite{DElia:2018xwo,Endrodi:2019zrl,Ali:2024mnn}.
Moreover, based on the observed strengthening of the chiral transition, it was conjectured to become first order
at strong magnetic fields~\cite{Endrodi:2015oba,Ding:2020inp}: numerical evidence for a first order line
was eventually shown in Ref.~\cite{DElia:2021yvk}, with a critical endpoint constrained in the range
$(T,~eB) = (63 \div 98$~MeV, $4 \div 9$~GeV$^2$). Many other non-trivial effects have been discussed in the literature:
Refs.~\cite{Astrakhantsev:2019zkr,Almirante:2024lqn} studied the electrical
conductivity of the quark-gluon plasma in the presence of a background magnetic field and found that the conductivity along the magnetic field
grows with the magnetic field; Ref.~\cite{Ding:2023bft} investigated the effect of an external magnetic field on the fluctuations and correlations
of conserved charges, which are used to probe the QCD phase structure in heavy-ion collision experiments; Ref.~\cite{Astrakhantsev:2024mat} investigated the QCD equation of state at finite density in an external magnetic field; 
Refs.~\cite{Bonati:2014ksa,Bonati:2016kxj,Bonati:2017uvz,DElia:2021tfb} investigated the effect of the magnetic field on the confining properties of QCD.

In this work we explore the effect of a background magnetic field on the phase diagram of QCD at imaginary chemical potentials, in particular we
study how the presence of a finite magnetic field affects the RW endpoint. The Roberge-Weiss and chiral transitions are somewhat related: not only
the RW endpoint is attached to (the analytic continuation of) the chiral transition line, but the Roberge-Weiss temperature and the chiral restoration
temperatures have been found to be very close to each other in the chiral limit~\cite{Bonati:2018fvg,Cuteri:2022vwk}. Therefore, it is conceivable that
a magnetic field could affect the RW transition similarly to the chiral transition, by lowering the critical temperature and turning the transition to
first order. Furthermore, the first order line found for large $eB$ in the $T - eB$ plane is reminiscent of the similar expected in the $T - \mu_B$ plane, so exploring the effects of a baryon chemical potential (even if imaginary) on the magnetic first order line
could better define the terms of this possible analogy.
 
\section{Numerical set-up}
We ran lattice simulations with $N_f = 2+1$ stout-staggered fermions at the physical point using the tree-level Symanzik improved gauge action. The partition function is
$$Z=\int{[DU]}\,e^{-S_{YM}}\prod_{f=u,d,s} \det{(M_{st}^f)}^\frac{1}{4}\mbox{ , }$$
where the gauge action and the staggered fermion matrix are
\begin{eqnarray*}
  S_{YM} &=& -\frac{\beta}{3} \, \sum_{i, \mu\ne\nu}\left(\frac{5}{6}W^{1\times1}_{i,\mu\nu}-\frac{1}{12}W^{1\times2}_{i,\mu\nu}\right) \\
      {(M^f_{st\ })}_{ij} &=& \hat m_f\delta_{i,j}+\sum_{\mu=1}^{4}\frac{\eta_{i;\mu}}{2}
\left(U^{(2)}_{i;\mu}\delta_{i, j-\hat{\mu}} - U^{(2)\dagger}_{i-\hat{\mu};\mu}\delta_{i,j+\hat{\mu}}\right)\mbox{.}
\end{eqnarray*}
with $W^{1\times1}_{i,\mu\nu}$ and $W^{1\times2}_{i,\mu\nu}$ being the $1\times1$ and $1\times2$ Wilson loops, $\hat m_f$ the bare quark masses, $\eta_{i;\mu}$ the staggered quark phases and $U^{(2)}_{i;\mu}$ the twice stout smeared link variables.
Bare parameters have been tuned in order to move on a line of constant physics, based on the determinations
of Refs.~\cite{Aoki:2009sc,Borsanyi:2010cj,Borsanyi:2013bia}.

A uniform background magnetic field along the $z$ direction has been introduced on the lattice by adding $U(1)$ phases to the link variables $U^{(2)}_{i;\mu} \mapsto u_{i;\mu}^f  U^{(2)}_{i;\mu}$, with

$$u_{i;y}^f=e^{i a^2 q_f B \,i_x}, \qquad {u_{i;x}^f\vert}_{i_x=N_x}=e^{-ia^2q_fN_xBi_y}, \qquad {u_{i;x}^f\vert}_{i_x\neq N_x} = u_{i;z}^f = u_{i;t}^f = 1 \mbox{.}$$
The magnetic field must obey the quantization condition $q_f B = \frac{2 \pi b_z}{a^2 N_x N_y}$, with $b_z$ an integer number, and to avoid cut-off effects one would like to have $2 b_z / N_x N_y \ll 1$.

Simulations have been performed on $N_t=6,8$ lattices of different volumes $N_s^3$. For each lattice we varied the temperature while keeping the chemical potential fixed on the RW line $\mu_q / T = i \pi$. The Roberge-Weiss temperature has been estimated as the inflection point of the imaginary part of the Polyakov loop (which acts as the order parameter for the transition) and the peak of its susceptibility,

$$L = \langle |Im P|\rangle \mbox{\phantom{ooooo}} \chi_L = N_t N_s^3 (\langle |Im P|^2 \rangle - \langle |Im P| \rangle^2) \mbox{.}$$

The simulations have been repeated for several magnetic fields $eB = \frac{6\pi b_z}{(a N_s)^2} = \frac{6 \pi b_z N_t^2}{N_s^2} T^2$. For some magnetic fields we also applied a finite-size scaling analysis to assess the nature of the transition.

\section{Numerical results for $N_t=6$}
\begin{figure}[h!]
  \centering
  \includegraphics[width=0.5\linewidth, clip]{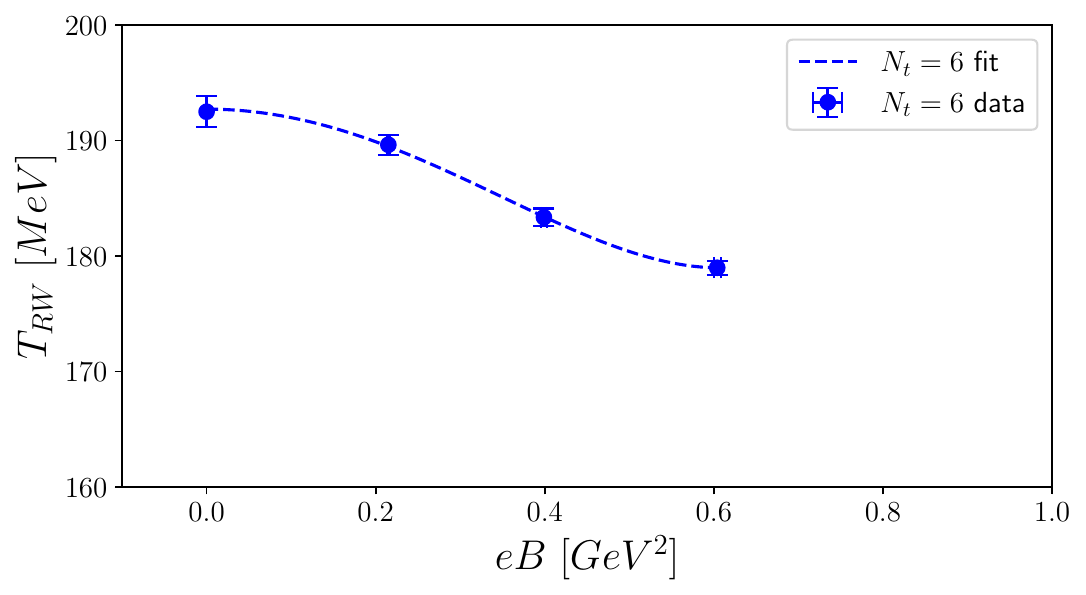}
  \caption{Roberge-Weiss temperature as a function of the magnetic field for $eB = 0.0$ to $0.6$~GeV$^2$. The blue error bars depict the numerical results from $N_t=6$ simulations, while the blue dashed line represents a fit using a rational function ansatz.}
  \label{fig:critline_loweb}
\end{figure}

We performed an initial series of runs at $eB = 0.2,~ 0.4$ and $0.6$~GeV$^2$ on $N_t=6$ lattices. Finite-size effects were checked by comparing the results for $N_s=18,~24$ lattices. These effects are small, for instance at $eB = 0.6$~GeV$^2$ we have calculated $T_{RW} = 180.38(69)$~MeV for $N_s=18$ and $T_{RW} = 178.99(59)$~MeV for $N_s=24$. The central values are in agreement within $1.5~\sigma$. The Roberge-Weiss temperature is found to decreases monotonically with  an increasing magnetic field. Numerical results are illustrated in Fig. \ref{fig:critline_loweb}, where we also show the results at $eB = 0$~GeV$^2$ from a previous work. The blue error bars denote the results from the simulations.  The dashed curve is the result of a fit using the rational function ansatz $T_{RW}(eB) = T_{RW}(0) \frac{1 + a(eB)^2}{1 + b(eB)^2}$. This is the same ansatz employed in Ref. \cite{Endrodi:2015oba} to parametrize the chiral transition line on the $(\mu_q=0, T, eB)$ plane, and it describes well our data, at least at these relatively low magnetic fields. We notice that the critical line changes curvature at $eB \sim 0.6$~GeV$^2$, close to the magnetic flipping point where Ref. \cite{Braguta:2019yci} observed a qualitative change in the behavior of the physical curvature of the chiral transition at nonzero baryon density. 

\begin{figure}[h!]
  \centering
  \includegraphics[width=1.0\linewidth, clip]{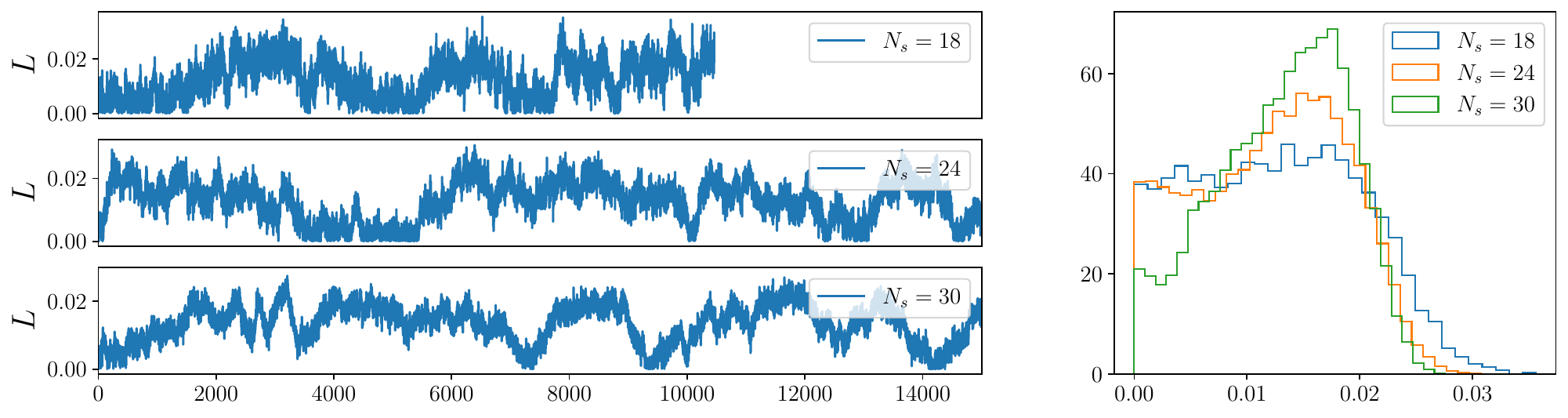}
  \caption{Monte Carlo histories (left panel) and histograms (right panel) of the Polyakov loop at $eB=1.0$~GeV$^2$ close to the critical temperature. Results from $N_t=6$ simulations with spatial extensions $N_s = 18, 24, 30$.}
  \label{fig:1p0GeV2_Nt6_hist}
\end{figure}

\begin{figure}[h!]
  \centering
  \includegraphics[width=0.497\linewidth, clip]{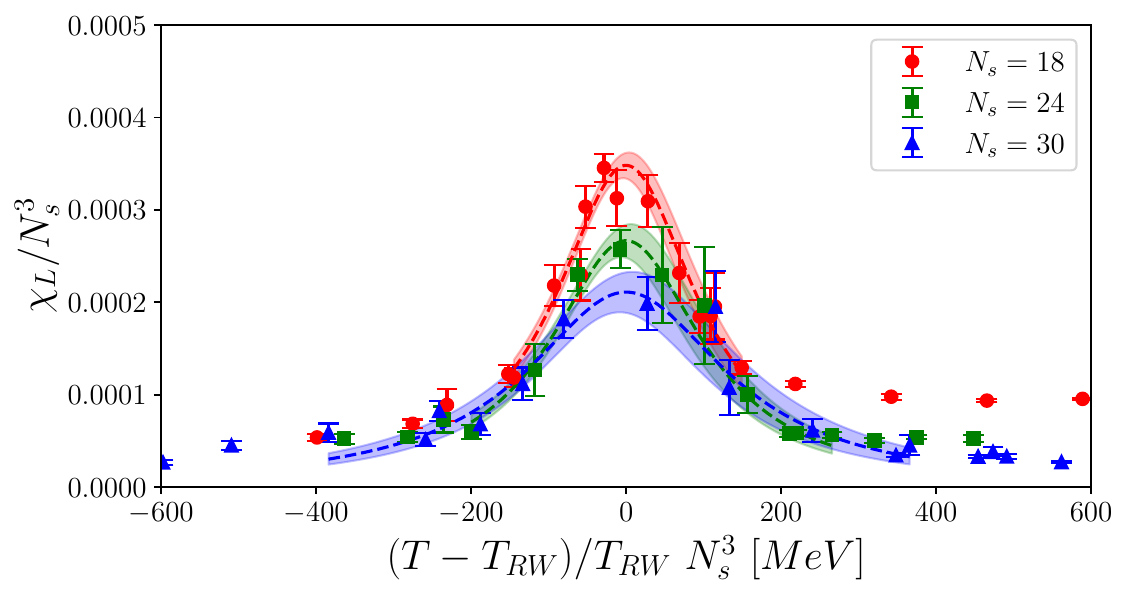}
  \includegraphics[width=0.497\linewidth, clip]{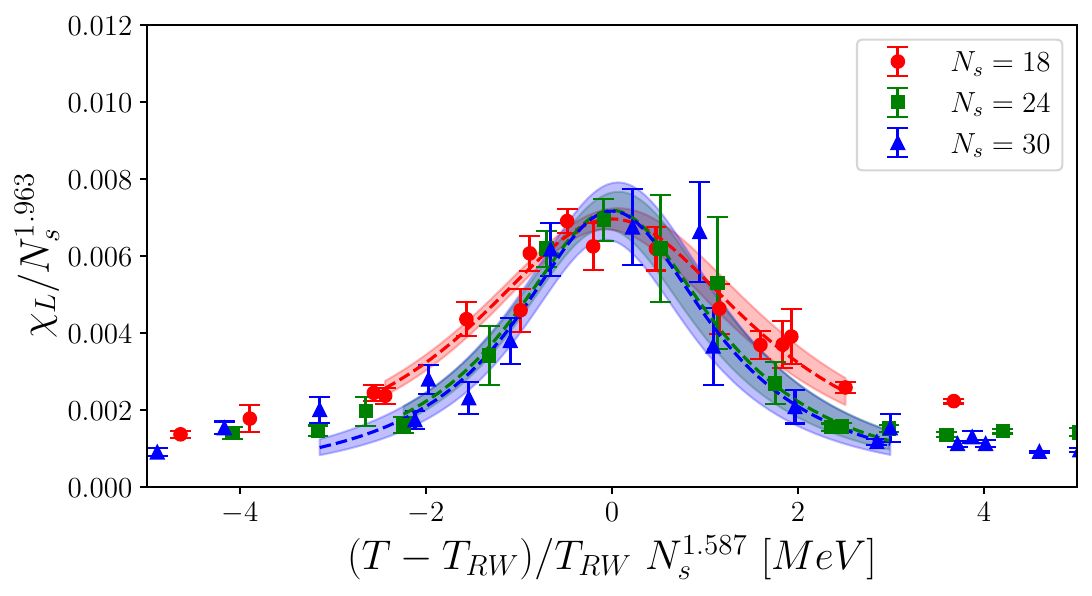}
  \caption{Collapse plots of the susceptibility at $eB=1.0$~GeV$^2$ for a first order (left panel) and second order $Z_2$ (right panel) transition. Results from $N_t=6$ simulations with spatial extensions $N_s = 18, 24, 30$.}
  \label{fig:1p0GeV2_Nt6_collapseplots}
\end{figure}

We also conducted additional runs in the strong magnetic field regime, for $eB = 1.0$ and $2.5$~GeV$^2$. For $eB = 1.0$~GeV$^2$ the simulations were performed for three spatial extensions $N_s=18,24,30$. Fig. \ref{fig:1p0GeV2_Nt6_hist} shows the Monte Carlo histories (left panel) and histograms (right panel) of the order parameter obtained for the three volumes close to $T_{RW}$. The absence of a double peaked structure in the histograms suggests that the transition has not yet turned first order. This has been confirmed by a finite-size scaling analysis. The susceptibility is expected to scale as
\begin{equation}
\chi_L = N_s^{\frac{\gamma}{\nu}} \phi (\tau N_s^\frac{1}{\nu})\mbox{ ,}
\label{eq:fss}
\end{equation}
where $\tau = \frac{T - T_{RW}}{T_{RW}}$ is the reduced temperature and $\gamma, \nu$ are the critical exponents. Consequently, when plotting $\chi_L  / N_s^{\frac{\gamma}{\nu}}$ as a function of $\tau N_s^\frac{1}{\nu}$, the plots for the different lattice sizes are expected to collapse onto each other. Moreover from Eq.~(\ref{eq:fss}) it also follows that the peaks of the susceptibility are expected to scale $\propto N_s^{\frac{\gamma}{\nu}}$, thus the ratio  $\frac{\gamma}{\nu}$ can be estimated by fitting the peaks to $\chi_L^{MAX}(N_s) = a~N_s^b$.
Fig. \ref{fig:1p0GeV2_Nt6_collapseplots} shows the collapse plots for a first order transition (left panel) and a second order transition of the $Z_2$ universality class (right panel). The data are compatible with a second order transition. A fit of the susceptibility peaks yields $\frac{\gamma}{\nu} = 2.04(19)$, compatible with the critical exponents of a second order transition of the $Z_2$ universality class.

\FloatBarrier

\begin{figure}[h!]
  \centering
  \includegraphics[width=1.0\linewidth, clip]{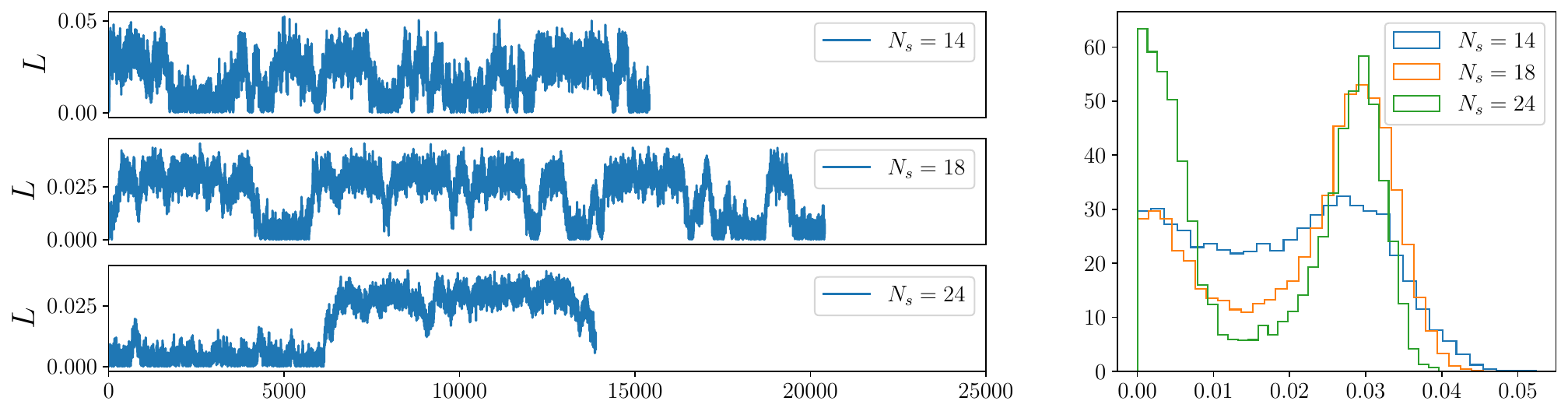}
  \caption{Monte Carlo histories (left panel) and histograms (right panel) of the Polyakov loop at $eB=2.5$~GeV$^2$ close to the critical temperature. Results from $N_t=6$ simulations with spatial extensions $N_s = 14, 18, 24$.}
  \label{fig:2p5GeV2_Nt6_hist}
\end{figure}

A qualitatively different picture emerges at $eB = 2.5$~GeV$^2$. This can be seen in Fig.~\ref{fig:2p5GeV2_Nt6_hist}, illustrating the Monte Carlo histories (left panel) and the histograms (right panel) of the Polyakov loop obtained close to the critical temperature for spatial extensions $N_s = 14, 18, 24$. The histograms have a double peaked distribution, which becomes more enhanced as the volume is increased, suggesting the presence of metastable states typical of a first order phase transition.

\FloatBarrier

\section{Numerical results for $N_t=8$}
The numerical results from $N_t=6$ simulations indicate that the transition is second order at $eB = 1.0$~GeV$^2$ and first order at $eB = 2.5$~GeV$^2$, suggesting the existence of a critical point between these two magnetic fields. The simulations at $eB = 1.0$~GeV$^2$ and $2.5$~GeV$^2$ have been repeated on $N_t=8$ lattices to confirm the stability of these conclusions as we approach the continuum limit. This is especially important at $eB = 2.5$~GeV$^2$, where large discretization effects can be expected, since for the $N_t=6$ lattices the magnetic field is not that far from the cut-off $eB = 2\pi / a^2 \sim 5$~GeV$^2$.

\begin{figure}[h!]
  \centering
  \includegraphics[width=0.497\linewidth, clip]{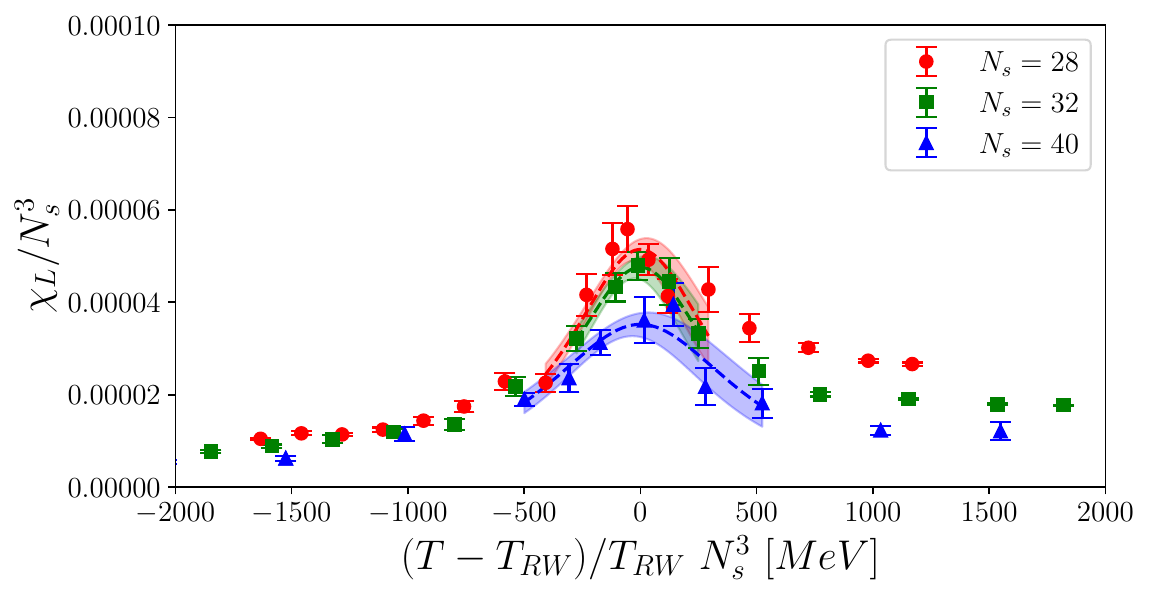}
  \includegraphics[width=0.497\linewidth, clip]{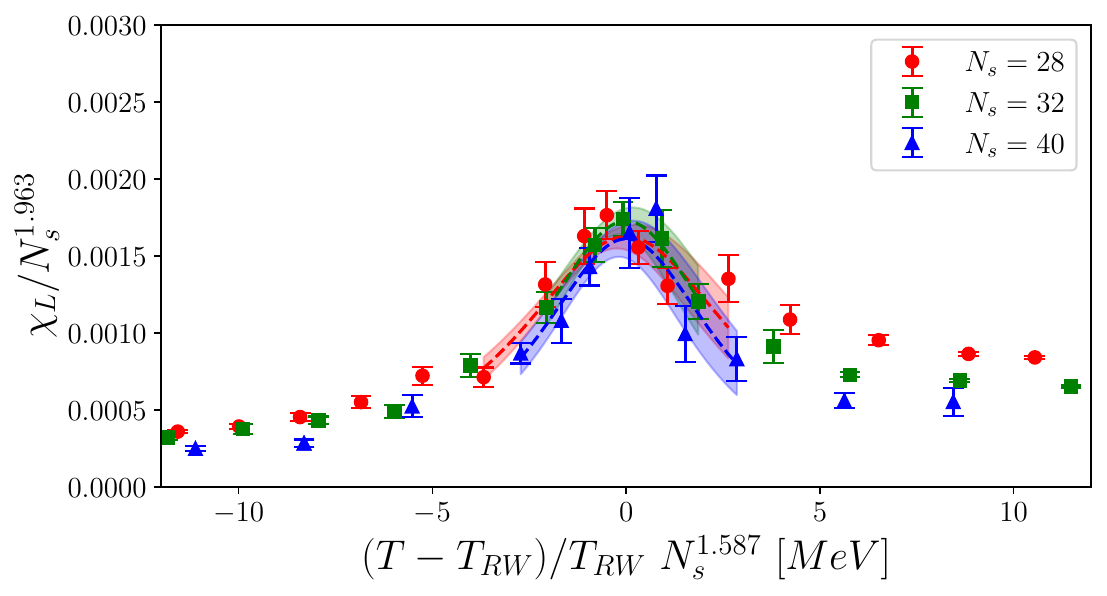}
  \caption{Collapse plots of the susceptibility at $eB=1.0$~GeV$^2$ for a first order (left panel) and second order $Z_2$ (right panel) transition. Results from $N_t=8$ simulations with spatial extensions $N_s = 28, 32, 40$.}
  \label{fig:1p0GeV2_Nt8_collapseplots}
\end{figure}

\begin{figure}[h!]
  \centering
  \includegraphics[width=1.0\linewidth, clip]{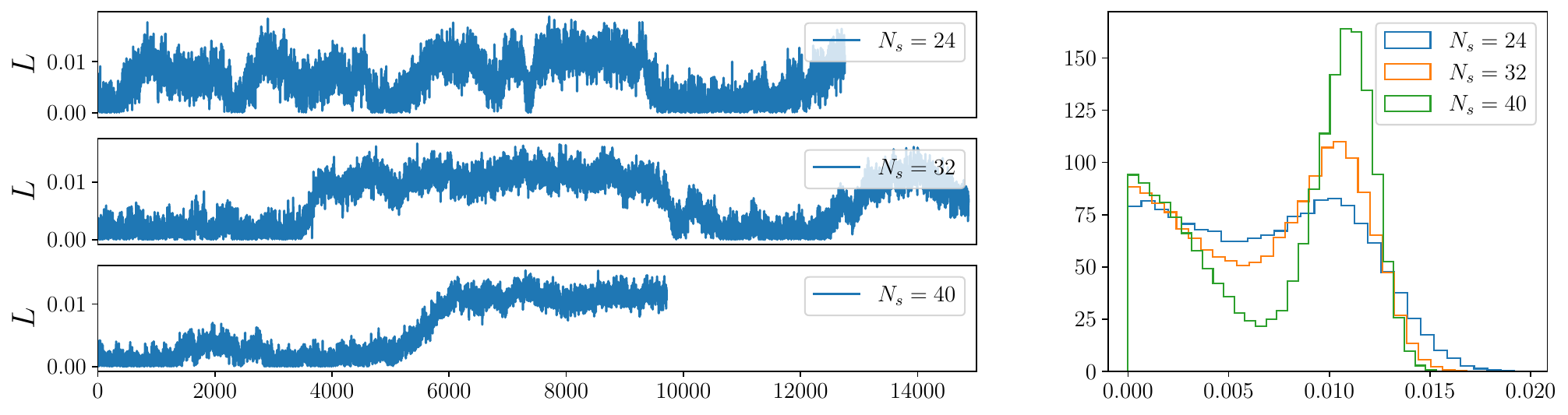}
  \caption{Monte Carlo histories (left panel) and histograms (right panel) of the Polyakov loop at $eB=2.5$~GeV$^2$ close to the critical temperature. Results from $N_t=8$ simulations with spatial extensions $N_s = 24, 32, 40$.}
  \label{fig:2p5GeV2_Nt8_hist}
\end{figure}

\begin{figure}[h!]
  \centering
  \includegraphics[width=0.497\linewidth, clip]{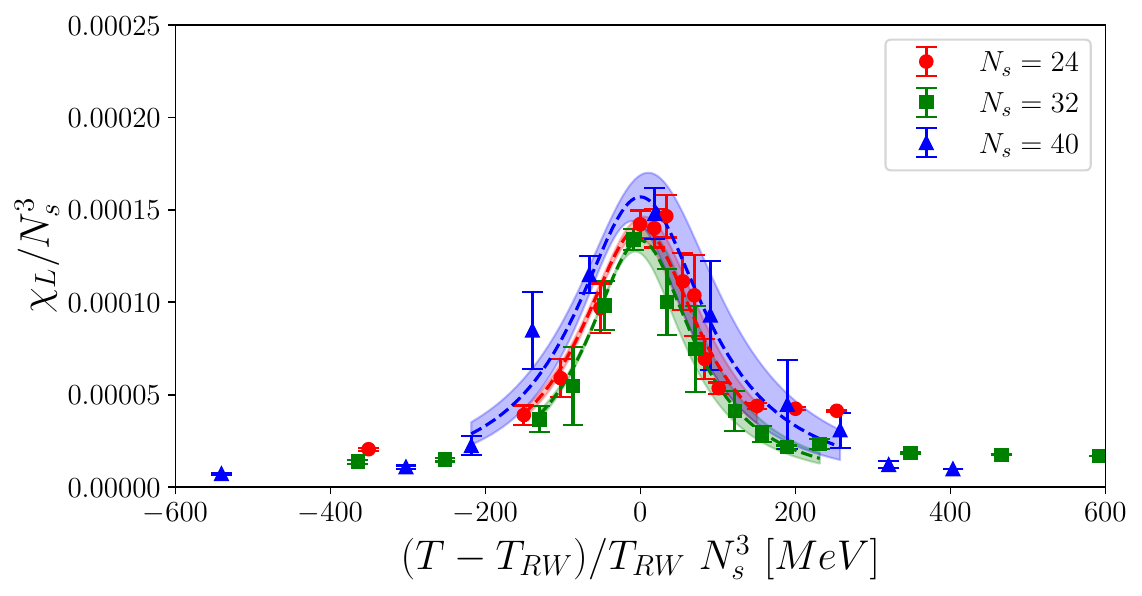}
  \includegraphics[width=0.497\linewidth, clip]{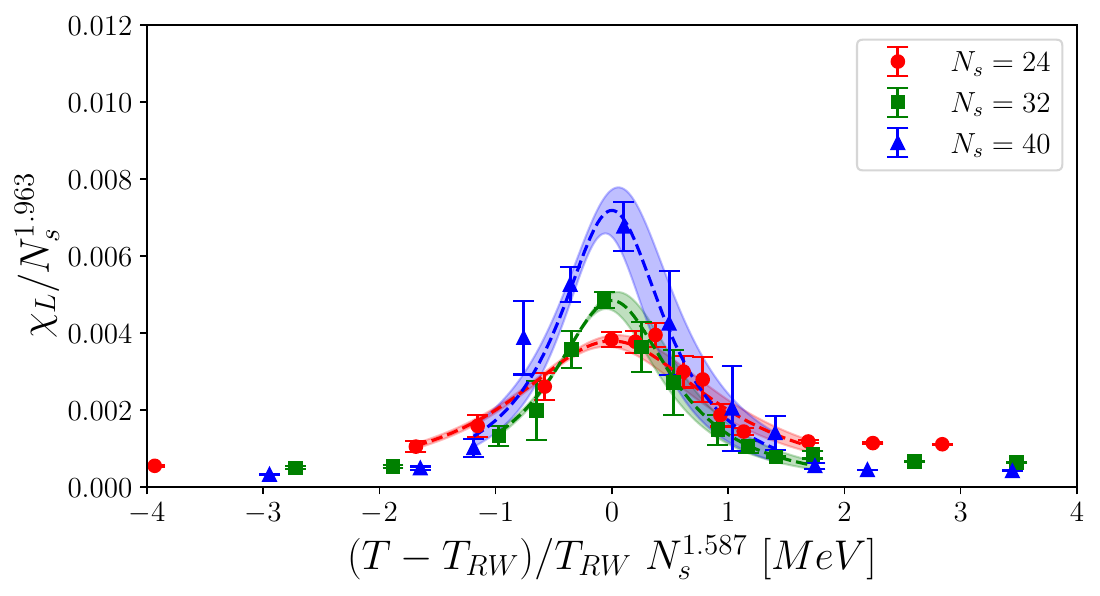}
  \caption{Collapse plots of the susceptibility at $eB=2.5$~GeV$^2$ for a first order (left panel) and second order $Z_2$ (right panel) transition. Results from $N_t=8$ simulations with spatial extensions $N_s = 24, 32, 40$.}
  \label{fig:2p5GeV2_Nt8_collapseplots}
\end{figure}

The results of the finite-size scaling analysis at $eB = 1.0$~GeV$^2$ are illustrated in the collapse plots shown in Fig. \ref{fig:1p0GeV2_Nt8_collapseplots} for both a first order (left panel) and second order (right panel) phase transition. Even on these finer lattices the RW transition is compatible with a second order transition of the $Z_2$ universality class. This is further supported by fitting the susceptibility peaks, which yields $\frac{\gamma}{\nu} = 1.97(28)$. Conversely, at $eB = 2.5$~GeV$^2$ the Monte Carlo histories and the histograms of the order parameter (see respectively the left and right panels of Fig. \ref{fig:2p5GeV2_Nt8_hist}) close to $T_{RW}$ still suggest a first order phase transition. This is confirmed by the collapse plots displayed in Fig. \ref{fig:2p5GeV2_Nt8_collapseplots} and by a fit of the susceptibility peaks, which yields $\frac{\gamma}{\nu} = 3.03(18)$.

\FloatBarrier

\section{Curvature of the critical line on the Roberge-Weiss plane}
\begin{figure}[h!]
  \centering
  \includegraphics[width=0.5\linewidth, clip]{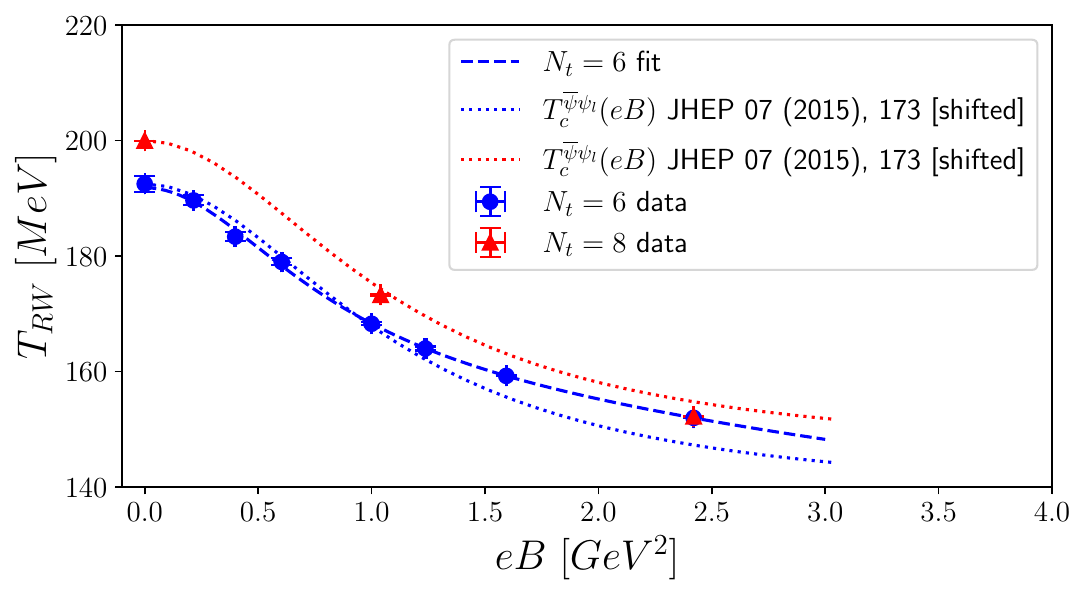}
  \caption{Roberge-Weiss temperature as a function of the magnetic field for $eB = 0.0$ to $2.5$~GeV$^2$. The blue error bars depict the numerical results from $N_t=6$ simulations, while the red error bars represent the numerical results from $N_t=8$ simulations. The dashed blue line represents a fit using a rational function ansatz. The dotted lines represent the parametrization obtained in Ref.~\cite{Endrodi:2015oba} for the chiral (pseudo)critical line on the $\mu_q = 0$ plane by studying the light quark condensate. The parametrization has been shifted along the temperature axis to match the Roberge-Weiss temperature at zero magnetic field for $N_t=6$ (blue) and $N_t =8$ (red).}
  \label{fig:critline_full}
\end{figure}

So far we have shown that the Roberge-Weiss temperature decreases monotonically with an increasing magnetic field and that the transition becomes first order somewhere between $1.0$~GeV$^2$ and $2.5$~GeV$^2$. Fig.~\ref{fig:critline_full} summarizes the estimates obtained for $T_{RW}$ in this work. The blue error bars represent the results from $N_t=6$ simulations, while the red error bars depict the results from $N_t=8$ simulations. A rational function ansatz $T^{(2)}_{RW}(eB) = T_{RW}(0) \frac{1 + a(eB)^2}{1 + b(eB)^2}$ fits well our $N_t=6$ data up to $1.6$~GeV$^2$. However an higher order rational ansatz is necessary to maintain a good quality of fit when including the RW temperature at $eB = 2.5$~GeV$^2$. The dashed blue line represents a fit done using the rational function ansatz $T^{(4)}_{RW}(eB) = T_{RW}(0) \frac{1 + a(eB)^2  + c (eB)^4}{1 + b(eB)^2  + d(eB)^4}$.

Additionally, Fig. \ref{fig:critline_full} shows as dotted lines the rational function parametrization obtained in Ref.~\cite{Endrodi:2015oba} for the chiral (pseudo)critical line on the $(\mu_q = 0, T, eB)$ plane using the light quark condensate as a probe for the chiral transition. The parametrization has been shifted along the temperature axis to match the Roberge-Weiss temperature at zero magnetic field for $N_t=6$ (blue) and $N_t =8$ (red). We notice that the curvature seemingly matches the curvature of our data. Indeed by Taylor expanding our rational function ansatze around $eB = 0$~GeV$^2$ we find the curvature coefficients $k=-50.0(3.5)$ and $k=-56(10)$, respectively for $T^{(2)}_{RW}(eB)$ and $T^{(4)}_{RW}(eB)$. These values both agree within errors with the curvature $k \sim -44.8$ of the parametrization found in Ref. \cite{Endrodi:2015oba}.

\FloatBarrier

\section{Conclusions}
In this work we have studied the Roberge-Weiss endpoint in the presence of a background magnetic field. Similarly to what happens for the chiral transition, a magnetic field reduces the Roberge-Weiss temperature and has the effect of turning the transition to first order at strong magnetic fields. This effect occurs somewhere between $eB=1.0$ and $eB=2.5$~GeV$^2$, as determined from $N_t=6$ lattice simulations and later confirmed using finer lattices. The critical line on the Roberge-Weiss plane can be parametrized by a rational function ansatz. Its curvature was estimated to be either $k=-50.0(3.5)$ and $k=-56(10)$, depending on the order of the rational function used for the parametrization. This curvature is similar to the curvature of the chiral (pseudo)critical line on the $\mu_q = 0$ plane estimated in the literature from the light quark condensate.

\acknowledgments
Numerical simulations have been performed on the Marconi 100 and Leonardo clusters at CINECA, based on the agreement between INFN and CINECA under projects INF23\_npqcd and INF24\_npqcd. KZ acknowledges support by the project “Non-perturbative aspects of fundamental interactions, in the Standard Model and beyond” funded by MUR, Progetti di Ricerca di Rilevante Interesse Nazionale (PRIN), Bando 2022, Grant 2022TJFCYB (CUP I53D23001440006).

\bibliographystyle{JHEP}
\bibliography{biblio}

\end{document}